\def\papertitle{A GENERAL-PURPOSE DEEP LEARNING APPROACH TO MODEL TIME-VARYING AUDIO EFFECTS}
\def\paperauthorA{Marco A. Mart\'{i}nez Ram\'{i}rez, Emmanouil Benetos, Joshua D. Reiss}

\documentclass[twoside,a4paper]{article}
\usepackage{paper_style}
\usepackage{amsmath,amssymb,amsfonts,amsthm}
\usepackage{euscript}
\usepackage[latin1]{inputenc}
\usepackage[T1]{fontenc}
\usepackage{ifpdf}

\usepackage[english]{babel}
\usepackage{caption}
\usepackage{subfig}
\usepackage{color}

\usepackage{booktabs}

\ninept

\usepackage{times}

\newif\ifpdf
\ifx\pdfoutput\relax
\else
   \ifcase\pdfoutput
      \pdffalse
   \else
      \pdftrue
\fi

\ifpdf
  \usepackage[pdftex,
    pdftitle={\papertitle},
    pdfauthor={\paperauthorA},
    colorlinks=false, 
    bookmarksnumbered, 
    pdfstartview=XYZ 
  ]{hyperref}
  \usepackage[pdftex]{graphicx}

\else 
  \usepackage[dvips]{epsfig,graphicx}
  \usepackage[dvips,
    colorlinks=false,
    bookmarksnumbered, 
    pdfstartview=XYZ
  ]{hyperref}
  
\fi
  \usepackage[hypcap=true]{caption}
\title{\papertitle}

\affiliation{
\paperauthorA }
{{Centre for Digital Music,} \\ Queen Mary University of London\\ London, United Kingdom\\
 {\tt \href{mailto:m.a.martinezramirez@qmul.ac.uk}{{m.a.martinezramirez, emmanouil.benetos, joshua.reiss}@qmul.ac.uk}}
}

\begin{document}
\ifpdf
  \DeclareGraphicsExtensions{.png,.jpg,.pdf}
\else 
  \DeclareGraphicsExtensions{.eps}
\fi

\maketitle

\graphicspath{{hres/}}

\begin{abstract}

Audio processors whose parameters are modified periodically over time are often referred as time-varying or modulation based audio effects. Most existing methods for modeling these type of effect units are often optimized to a very specific circuit and cannot be efficiently generalized to other time-varying effects. Based on convolutional and recurrent neural networks, we propose a deep learning architecture for generic black-box modeling of audio processors with long-term memory. We explore the capabilities of deep neural networks to learn such long temporal dependencies and we show the network modeling various linear and nonlinear, time-varying and time-invariant audio effects. In order to measure the performance of the model, we propose an objective metric based on the psychoacoustics of modulation frequency perception. We also analyze what the model is actually learning and how the given task is accomplished.

\end{abstract}

\section{Introduction}
\label{sec:intro}

Modulation based or time-varying audio effects involve audio processors or effect units that include a modulator signal within their analog or digital implementation \cite{reiss2014audio}. These modulator signals are in the low frequency range (usually below $20$ Hz). Their waveforms are based on common periodic signals such as sinusoidal, squarewave or sawtooth oscillators and are often referred to as a Low Frequency Oscillator (LFO). The LFO periodically modulates certain parameters of the audio processors to alter the timbre, frequency, loudness or spatialization characteristics of the audio. This differs from time-invariant audio effects which do not change their behavior over time. Based on how the LFO is employed and the underlying signal processing techniques used when designing the effect units, we can classify modulation based audio effects into \textit{time-varying filters} such as phaser or wah-wah; \textit{delay-line based} effects such as flanger or chorus; and \textit{amplitude modulation} effects such as tremolo or ring modulator \cite{zolzer2011dafx}. 

The \textit{phaser} effect is a type of time-varying filter implemented through a cascade of notch or all-pass filters. The characteristic sweeping sound of this effect is obtained by modulating the center frequency of the filters, which creates phase cancellations or enhancements when combining the filter's output with the input audio. Similarly,  the \textit{wah-wah} is based on a bandpass filter with a variable center frequency, usually controlled by a pedal. If the center frequency is modulated by an LFO or an envelope follower, the effect is commonly called \textit{auto-wah}.

Delay-line based audio effects, as in the case of \textit{flanger} and \textit{chorus}, are based on the modulation of the length of the delay lines. A \textit{flanger} is implemented via a modulated comb filter whose output is mixed with the input audio. Unlike the phaser, the notch and peak frequencies caused by the flanger's sweep comb filter effect are equally spaced in the spectrum, thus causing the known metallic sound associated with this effect. A \textit{chorus} occurs when mixing the input audio with delayed and pitch modulated copies of the original signal. This is similar to various musical sources playing the same instrument but slightly shifted in time. \textit{vibrato} is digitally implemented as a delay-line based audio effect, where pitch shifting is achieved when periodically varying the delay time of the input waveform \cite{smith2010physical}.

\textit{Tremolo} is an amplitude modulation effect where an LFO is used to directly vary the amplitude of the incoming audio, creating in this way a perceptual temporal fluctuation. A \textit{ring modulator} is also based on amplitude modulation, but the modulation is achieved by having the input audio multiplied by a sinusoidal oscillator with higher carrier frequencies. In the analog domain, this effect is commonly implemented with a diode bridge, which adds a nonlinear behavior and a distinct sound to this effect unit. Another type of modulation based effect that combines amplitude, pitch and spatial modulation is the \textit{Leslie speaker}, which is implemented by a rotating horn and a rotating woofer inside a wooden cabinet. This effect can be interpreted as a combination of \textit{tremolo}, Doppler effect and reverberation \cite{henricksen1981unearthing}. 

Most of these effects can be implemented directly in the digital domain through the use of digital filters and delay lines. Nevertheless, modeling specific effect units or analog circuits has been heavily researched and remains an active field. This is because hardware effect units are characterized by the nonlinearities introduced by certain circuit components. Musicians often prefer the analog counterparts because the digital implementations may lack this behavior, or because the digital simulations make certain assumptions when modeling specific nonlinearities. 

Virtual analog methods for modeling such effect units mainly involve circuit modeling and optimization for specific analog components such as operational amplifiers or transistors. This often requires assumptions or models that are too specific for a certain circuit. Such models are also not easily transferable to different effects units since expert knowledge of the type of circuit being modeled is required, i.e. specific linear and nonlinear components. 

Prior to this work, deep learning architectures have not yet been implemented to model time-varying audio effects. Thus, building on \cite{martinez2018end,martinez2019modeling}, we propose a general-purpose deep learning approach to model this type of audio effects. Using convolutional, recurrent and fully connected layers, we explore how a deep neural network (DNN) can learn the long temporal dependencies which characterizes these effect units as well as the possibilities to match nonlinearities within the audio effects. We include Bidirectional Long Short-Term Memory (Bi-LSTM) neural networks and explore their capabilities when learning time-varying transformations. We explore linear and nonlinear time-varying emulation as a content-based transformation without explicitly obtaining the solution of the time-varying system. 

We show the model matching modulation based audio effects such as \textit{chorus}, \textit{flanger}, \textit{phaser}, \textit{tremolo}, \textit{vibrato}, \textit{tremolo-wah}, \textit{ring modulator} and \textit{Leslie speaker}. We investigate the capabilities of the model when adding further nonlinearities to the linear time-varying audio effects. Furthermore, we extend the applications of the model by including nonlinear time-invariant audio effects with long temporal dependencies such as \textit{auto-wah}, \textit{compressor} and \textit{multiband compressor}. Finally, we measure performance of the model using a metric based on the modulation spectrum.

The paper is structured as follows. In Section \ref{sec:background} we present the relevant literature related to virtual analog of modulation based audio effects. Section \ref{sec:methods} gives details of our model, the modulation based effect tasks and the proposed evaluation metric. Sections \ref{sec:analysis} and \ref{sec:conclusion} show the analysis, obtained results, and the respective conclusion. 

\section{Background}
\label{sec:background}

\subsection{Virtual analog modeling of time-varying audio effects}

Virtual analog audio effects aim to simulate an effect unit and recreate the sound of an analog reference circuit. Much of the active research models nonlinear audio processors such as distortion effects, compressors, amplifiers or vacuum tubes \cite{pakarinen2009review, giannoulis2012digital, yeh2010automated}. With respect to modeling time-varying audio effects, most of the research has been applied to develop white-box methods, i.e. in order to model the effect unit a complete study of the internal circuit is carried out. These methods use circuit simulation techniques to characterize various analog components such as diodes, transistors, operational amplifiers or integrated circuits.

In \cite{huovilainen2005enhanced}, \textit{phasers} implemented via Junction Field Effect Transistors (JFET) and Operational Transconductance Amplifiers (OTA) were modeled using circuit simulation techniques that discretize the differential equations that describe these components. Using a similar circuit modeling procedure, delay-line based effects are modeled, such as \textit{flanger} and \textit{chorus} as implemented with Bucket Brigade Delay (BBD) chips. BBD circuits have been widely used in analog delay-line based effect units and several digital emulations have been investigated. \cite{raffel2010practical} emulated BBD devices through circuit analysis and electrical measurements of the linear and nonlinear elements of the integrated circuit. \cite{holters2018combined} modeled BBDs as delay-lines with fixed length but variable sample rate. 

Based on BBD circuitry, a \textit{flanger} effect was modeled in \cite{mavcak2016simulation} via the nodal DK-method. This is a common method in virtual analog modeling \cite{yeh2012automated} where nonlinear filters are derived from the differential equations that describe an electrical circuit. In \cite{holters2011physical}, a \textit{wah-wah} pedal is implemented using the nodal DK-method and the method is extended to model the temporal fluctuations introduced by the continuous change of the pedal. In \cite{eichas2014physical}, the \textit{MXR Phase 90} phaser effect is modeled via a thorough circuit analysis and the DK-method. This effect unit is based on JFETs, and voltage and current measurements were performed to obtain the nonlinear characteristics of the transistors. 

Amplitude modulation effects such as an analog \textit{ring modulator} were modeled in \cite{parker2011simple}, where the diode bridge is emulated as a network of static nonlinearities. \cite{smith2002doppler} modeled the rotating horn of the \textit{Leslie speaker} via varying delay-lines, artificial reverberation and physical measurements from the rotating loudspeaker. \cite{pekonen2011computationally} also modeled the \textit{Leslie speaker} and achieved frequency modulation through time-varying spectral delay filters and amplitude modulation using a modulator signal. In both \textit{Leslie speaker} emulations, various physical characteristics of the effect are not taken into account, such as the frequency-dependent directivity of the loudspeaker and the effect of the wooden cabinet.

In \cite{kiiski2016timevariant}, gray-box modeling was proposed for linear time-varying audio effects. This differs from white-box modeling, since the method was based on input-output measurements but the time-varying filters were based on knowledge of analog \textit{phasers}. In this way, \textit{phaser} emulation was achieved by multiple measurements of the impulse response of a cascade of all-pass filters. 

Another method to model time-varying audio effects is discretizing electrical circuit elements via Wave Digital Filters (WDF). The Hammond organ vibrato/chorus was modeled using WDFs in \cite{werner2016computational}, and  \cite{bogason2017modeling} performed circuit modeling through WDFs to emulate modulation based effects that use OTAs.

\setcounter{figure}{0}
\begin{figure*}[ht]
\hspace*{-0.75cm} 
\centering
\makebox[0pt]{\includegraphics[width=1.1\linewidth]{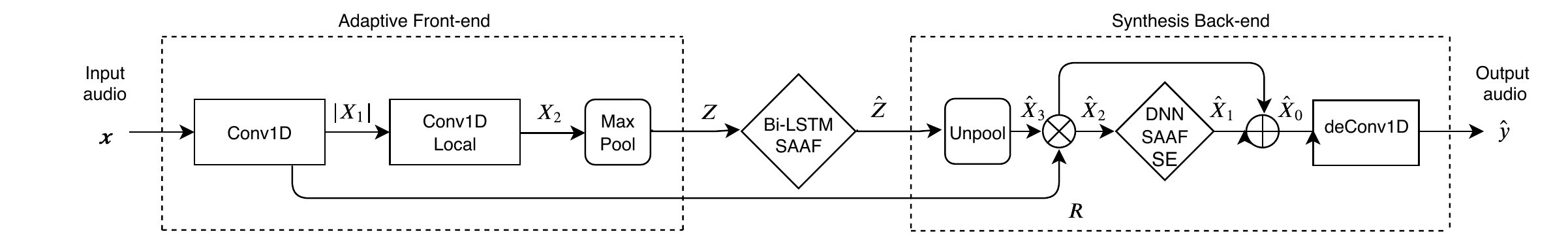}}
\caption{\label{fig:model}{\textit{Block diagram of the proposed model; adaptive front-end, Bi-LSTM and synthesis back-end.}}}
\end{figure*}

\subsection{End-to-end deep neural networks}

End-to-end deep learning is based on the idea that an entire problem can be taken as a single indivisible task which must be learned from input to output. Deep learning architectures using this principle have recently been researched in the music information retrieval field \cite{pons2017end, venkataramani2017adaptive, stoller2018wave}, since the amount of required prior knowledge may be reduced and engineering effort minimized by learning directly from raw audio \cite{dieleman2014end}. Recent work also demonstrated the feasibility of these architectures for audio synthesis and audio effects modeling. \cite{mehri2016samplernn,engel2017neural} proposed models that synthesize audio waveforms and \cite{blaauw2017neural} obtained a model capable of performing singing voice synthesis. 

End-to-end deep neural networks for audio effects modeling were implemented in \cite{martinez2018end}, where Equalization (EQ) matching was achieved with convolutional neural networks (CNN). Also, \cite{martinez2019modeling} presented a deep learning architecture for modeling nonlinear processors such as distortion, overdrive and amplifier emulation. The DNN is capable of modeling an arbitrary combination of linear and nonlinear memoryless audio effects, but does not generalize to transformations with long temporal dependencies such as modulation based audio effects.

\section{Methods}
\label{sec:methods}
\subsection{Model}

The model is entirely based on the time-domain and operates with raw audio as the input and processed audio as the output. It is divided into three parts: adaptive front-end, latent-space and synthesis back-end. A block diagram can be seen in Fig. \ref{fig:model} and its structure is described in detail in Table \ref{table:architecture}. We build on the architecture from \cite{martinez2019modeling}, since we incorporate Bi-LSTMs into the latent-space and we modify the structure of the synthesis back-end in order to allow the model to learn nonlinear time-varying transformations. 

\subsection{Adaptive front-end}

The front-end performs time-domain convolutions with the incoming audio. It follows a filter bank architecture and is designed to learn a latent representation for each audio effect modeling task. It consists of a convolutional encoder which contains two CNN layers, one pooling layer and one residual connection. This residual connection is used by the back-end to facilitate the synthesis of the waveform based on the specific time-varying transformation.

In order to allow the model to learn long-term memory dependencies, the input consists of the current audio frame $x(t)$ concatenated with the $k$ previous and $k$ subsequent frames. These frames are of size $N$ and sampled with a hop size $\tau$. The input $\boldsymbol{x}$ is described by (\ref{eq1:input}).

\begin{table}
\caption{\textit{Detailed architecture of a model with input frame size of $4096$ samples and $\pm4$ context frames. }}
\centering
\resizebox{0.75\columnwidth}{!}{
\renewcommand{\arraystretch}{1.0}
  \begin{tabular}{cccc}
  \toprule 
  Layer & Output shape & Units & Output \\ [1ex]
  \midrule 
  Input & (9, 4096, 1) & . & $\boldsymbol{x}$\\
  Conv1D  & (9, 4096, 32) & 32(64) & $X_1$\\
  Residual & (4096, 32) &. & $R$\\
  Abs     & (9, 4096, 32) &. &. \\
  Conv1D-Local  & (9, 4096, 32) & 32(128) &. \\
  Softplus     & (9, 4096, 32) &. &  $X_2$\\
  MaxPooling & (9, 64, 32) &. & $Z$\\
  \midrule
  Bi-LSTM & (64, 128) & 64 &. \\
  Bi-LSTM & (64, 64) & 32 &. \\
  Bi-LSTM & (64, 32) & 16 &. \\
  SAAF & (64, 32) & 25 & $\hat{Z}$\\
  \midrule
  Unpooling & (4096, 32) &. & $\hat{X}_3$\\
  Multiply  & (4096, 32) &. & $\hat{X}_2$\\
  Dense & (4096, 32) & 32 & .\\
  Dense & (4096, 16) & 16 &. \\
  Dense & (4096, 16) & 16 &. \\
  Dense & (4096, 32) & 32 &. \\
  SAAF & (4096, 32) & 25 &. \\
  Abs & (4096, 32) & .&. \\
  Global Average & (1, 32) &. &.\\
  Dense & (1, 512) & 512 &. \\
  Dense & (1, 32) & 32 & .\\
  Multiply & (4096, 32) &. &$\hat{X}_1$\\
  Add & (4096, 32) &. & $\hat{X}_0$\\
  deConv1D & (4096, 1) &. & $\hat{y}$\\
  \bottomrule   
  \end{tabular}
  }
\label{table:architecture}
\end{table}

\begin{equation}
    \boldsymbol{x} = x(t + j\tau), j = -k,...,k 
\label{eq1:input}
\end{equation}

The first convolutional layer has $32$ one-dimensional filters of size $64$ and is followed by the \textit{absolute value} as nonlinear activation function. The operation performed by the first layer can be described by (\ref{eq:2-firstLayer}).

\begin{equation}
\boldsymbol{X}_{1} = \boldsymbol{x} * \boldsymbol{W}_{1}
\label{eq:2-firstLayer}
\end{equation}

Where $\boldsymbol{X}_{1}$ is the feature map after the input audio $\boldsymbol{x}$ is convolved with the kernel matrix $\boldsymbol{W}_{1}$. $\boldsymbol{R}$ is the corresponding row in $\boldsymbol{X}_{1}$ for the frequency band decomposition of the current input frame $x(t)$. The back-end does not directly receive information from the past and subsequent context frames. 
The second layer has $32$ filters of size $128$ and each filter is locally connected. We follow a filter bank architecture since each filter is only applied to its corresponding row in $|\boldsymbol{X}_{1}|$ and so we significantly decrease the number of trainable parameters. This layer is followed by the \textit{softplus} nonlinearity \cite{glorot2011deep}, described by (\ref{eq:3-secondLayer}).

\begin{equation}
\boldsymbol{X}_{2} = \textit{softplus}(|\boldsymbol{X}_{1}| * \boldsymbol{W}_{2})
\label{eq:3-secondLayer}
\end{equation}

Where $\boldsymbol{X}_{2}$ is the second feature map obtained after the local convolution with $\boldsymbol{W}_{2}$, the kernel matrix of the second layer. The \textit{max-pooling} operation is a moving window of size $N/64$ applied over $\boldsymbol{X}_{2}$, where the maximum value within each window corresponds to the output. 

By using the \textit{absolute value} as activation function of the first layer and by having larger filters $\boldsymbol{W}_{2}$, we expect the front-end to learn smoother representations of the incoming audio, such as envelopes \cite{venkataramani2017adaptive}. All convolutions and pooling operations are time distributed, i.e the same convolution or pooling operation is applied to each of the $2\cdot k+1$ input frames. 

\subsection{Bidirectional LSTMs}

The latent-space consists of three Bi-LSTM layers of $64$, $32$, and $16$ units respectively. Bi-LSTMs are a type of recurrent neural network that can access long-term context from both backward and forward directions \cite{graves2013speech}. Bi-LSTMs are capable of learning long temporal dependencies when processing timeseries where the context of the input is needed \cite{graves2005framewise}. 

The Bi-LSTMs process the latent-space representation $Z$. $Z$ is learned by the front-end and contains information regarding the $2\cdot k+1$ input frames. These recurrent layers are trained to reduce the dimension of $Z$, while also learning a nonlinear modulation $\hat{Z}$. This new latent representation is fed into the synthesis back-end in order to reconstruct an audio signal that matches the time-varying task. Each Bi-LSTM has dropout and recurrent dropout rates of $0.1$ and the first two layers have the \textit{hyperbolic tangent} as activation function. 

The performance of CNNs in regression tasks has improved by using adaptive activation functions  \cite{hou2017convnets}. So we add a Smooth Adaptive Activation Function (SAAF) as the nonlinearity for the last layer. SAAFs consist of piecewise second order polynomials which can approximate any continuous function and are regularized under a Lipschitz constant to ensure smoothness. As shown in \cite{martinez2019modeling}, SAAFs can be used within deep neural networks to model nonlinearities in audio processing tasks.

\subsection{Synthesis back-end}

The synthesis back-end accomplishes the reconstruction of the target audio by processing the current input frame $x(t)$ and the nonlinear modulation $\hat{Z}$. The back-end consists of an unpooling layer, a DNN block with SAAF and Squeeze-and-Excitation (SE) \cite{hu2018squeeze} layers (DNN-SAAF-SE) and a final CNN layer.

The DNN-SAAF-SE block consists of $4$ fully connected (FC) layers of $32$, $16$, $16$ and $32$ hidden units respectively. Each FC layer is followed by the \textit{hiperbolic tangent} function except for the last one, which is followed by a SAAF layer. Overall, each SAAF is locally connected and each function consists of $25$ intervals between $-1$ to $1$. 

The SE blocks explicitly model interdependencies between channels by adaptively scaling the channel-wise information of feature maps \cite{hu2018squeeze}. The SE dynamically scales each of the 32 channels and follows the structure from \cite{kim2018sample}. It consists of a global average pooling operation followed by two FC layers of $512$ and $32$ hidden units respectively. The FC layers are followed by a rectifier linear unit (\textit{ReLU}) and \textit{sigmoid} activation functions accordingly. Since the feature maps of the model are based on time-domain waveforms, we incorporate an \textit{absolute value} layer before the global average pooling operation.

The back-end matches the time-varying task by the following steps. First, a discrete approximation of $\boldsymbol{Z}$ ($\boldsymbol{\hat{X}}_{3}$) is obtained by an upsampling operation. Then the feature map $\boldsymbol{\hat{X}}_{2}$ is the result the element-wise multiplication of the residual connection $\boldsymbol{R}$ and $\boldsymbol{\hat{X}}_{3}$. This can be seen as a frequency dependent amplitude modulation between the learned modulator $\boldsymbol{Z}$ and the frequency band decomposition $\boldsymbol{R}$. 

\begin{equation}
\boldsymbol{\hat{X}_{2}} = \boldsymbol{\hat{X}}_{3} \cdot \boldsymbol{R}
\label{eq:4-modulation}
\end{equation}

The feature map $\boldsymbol{\hat{X}}_{1}$ is obtained when the nonlinear and channel-wise scaled filters from the DNN-SAAF-SE block are applied to the modulated frequency band decomposition $\boldsymbol{\hat{X}}_{2}$. Then, $\boldsymbol{\hat{X}}_{1}$ is added back to $\boldsymbol{\hat{X}}_{2}$, acting as a nonlinear delay-line.  

\begin{equation}
\boldsymbol{\hat{X}_{0}} = \boldsymbol{\hat{X}}_{1} + \boldsymbol{\hat{X}}_{2}
\label{eq:5-addition}
\end{equation}

The last layer corresponds to the deconvolution operation, which can be implemented by transposing the first layer transform. This layer is not trainable since its kernels are transposed versions of $\boldsymbol{W}_{1}$. In this way, the back-end reconstructs the audio waveform in the same manner that the front-end decomposed it. The complete waveform is synthesized using a \textit{hanning} window and constant overlap-add gain. 

All convolutions are along the time dimension and all strides are of unit value. The models have approximately $300k$ trainable parameters, which, within a deep learning context, represents a model that is not very large or difficult to train. 

\begin{figure}[b]
\center
\includegraphics[width=1.0\linewidth]{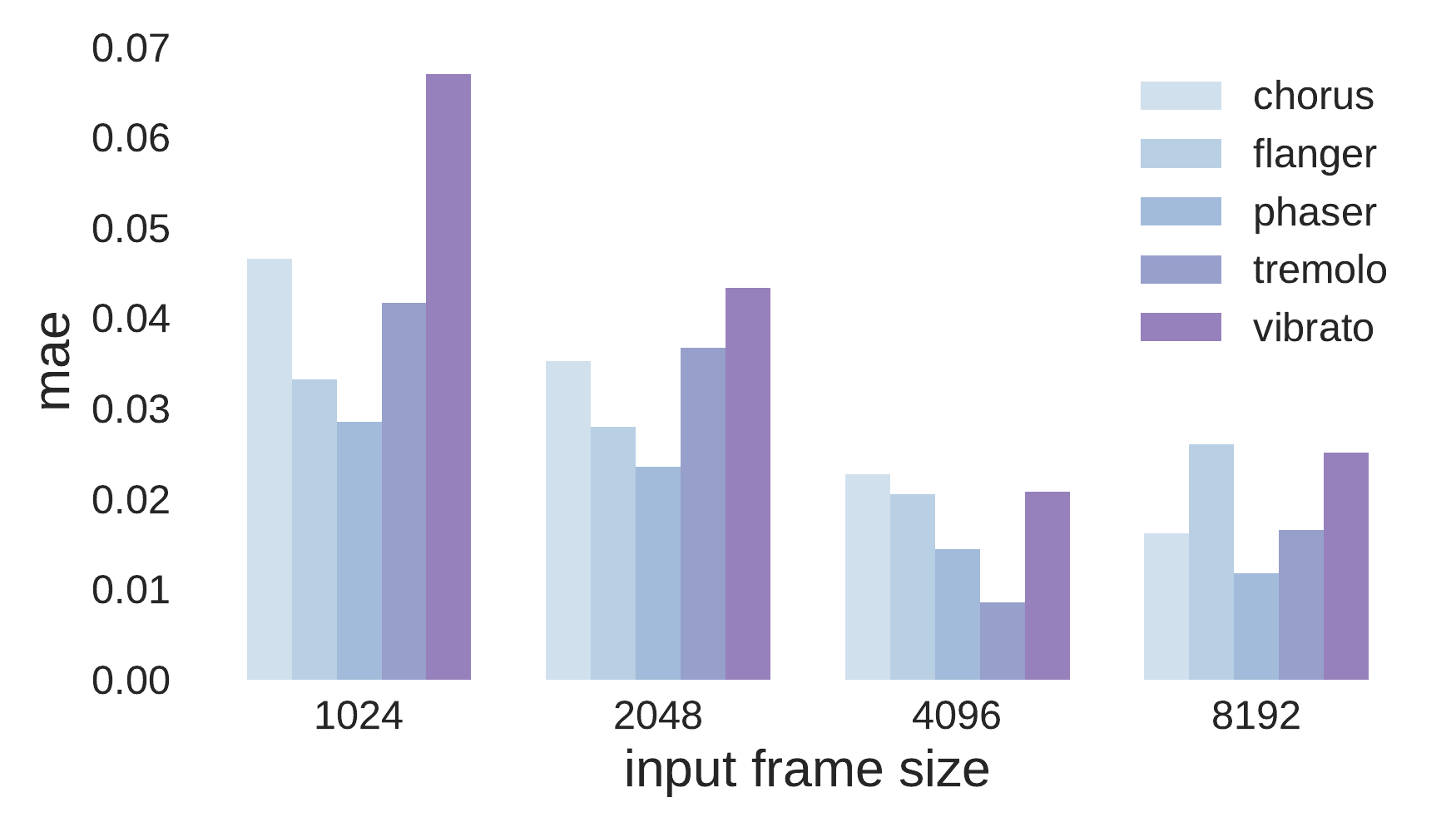}\\
\
\caption{\label{fig:mae-winlength}{\textit{ \textbf{mae} values for linear time-varying tasks with different input size frames.}}}
\end{figure}

\subsection{Training}

The training of the model is performed in two steps. The first step is to train only the convolutional layers for an unsupervised learning task, while the second step consists of an end-to-end supervised learning task based on a given time-varying target. During the first step only the weights of \textit{Conv1D} and \textit{Conv1D-Local} are trained and both the raw audio $x(t)$ and wet audio $y(t)$ are used as input and target functions.  This means the model is being prepared to reconstruct the input and target data in order to have a better fitting when training for the time-varying task. Only during this step, the unpooling layer of the back-end uses the time positions of the maximum values recorded by the \textit{max-pooling} operation.

Once the model is pretrained, the Bi-LSTM and DNN-SAAF-SE layers are incorporated into the model, and all the weights of the convolutional, recurrent, dense and activation layers are updated. Since small amplitude errors are as important as large ones, the loss function to be minimized is the mean absolute error between the target and output waveforms. We explore input size frames from $1024$ to $8192$ samples and we always use a hop size of $50\%$. The batch size consisted of the total number of frames per audio sample. 

\textit{Adam} is used as optimizer and we perform the pretraining for $200$ epochs and the supervised training for $500$ epochs. In order to speed convergence, during the second training step we start with a learning rate of $5\cdot 10^{-5}$ and we reduce it by $50\%$ every $150$ epochs. We select the model with the lowest error for the validation subset. 

\begin{figure*}[t]
\center
\includegraphics[width=0.9\linewidth]{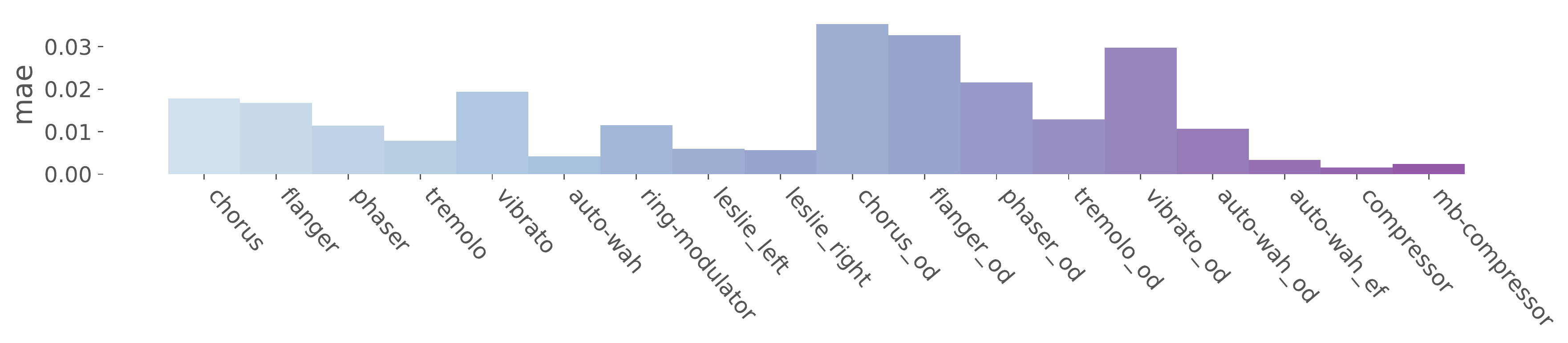}\label{fig:mae-error}\\ 
\includegraphics[width=0.9\linewidth]{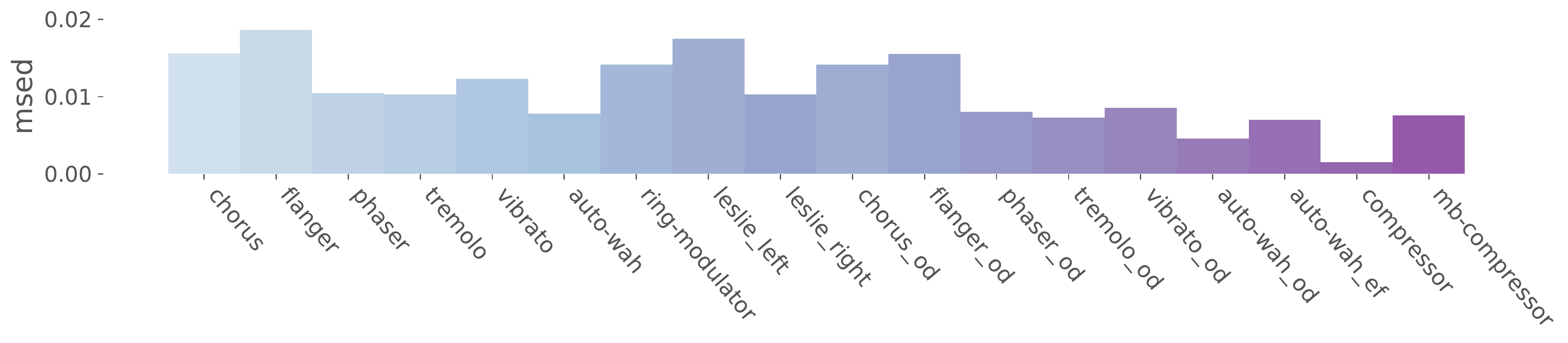}\label{fig:msed-error}\\
\caption{\textit{ \textbf{mae} and \textbf{msed} values with the test dataset for all the time-varying tasks. \emph{od}, \emph{ef} and \emph{mb} mean \textit{overdrive}, \textit{envelope follower} and \textit{multiband} respectively.}}
\label{fig:errors}
\end{figure*}
\subsection{Dataset}

Modulation based audio effects such as \textit{chorus}, \textit{flanger}, \textit{phaser}, \textit{tremolo} and \textit{vibrato} were obtained from the \textit{IDMT-SMT-Audio-Effects} dataset \cite{stein2010automatic}. It corresponds to individual 2-second notes and covers the common pitch range of various 6-string electric guitars and 4-string bass guitars. 

The recordings include the raw notes and their respective effected versions for $3$ different settings for each effect. For our experiments, for each of the above effects, we only use the setting $\#2$ from where we obtained the unprocessed and processed audio for bass guitar. In addition, processing the bass guitar raw audio, we implemented an \textit{auto-wah} with a peak filter whose center frequency ranges from $500$ Hz to $3$ kHz and modulated by a $5$ Hz sinusoidal. 
 
Since the previous audio effects are linear time-varying, we further test the capabilities of the model by adding a nonlinearity to each of these effects. Thus, using the bass guitar wet audio, we applied an \textit{overdrive} (gain$=+10$dB) after each modulation based effect.

We also use virtual analog implementations of a \textit{ring modulator} and a \textit{Leslie speaker} to process the electric guitar raw audio. The \textit{ring modulator} implementation\footnote{https://github.com/nrlakin/robot\_voice/blob/master/robot.py} is based on \cite{parker2011simple} and we use a modulator signal of $5$ Hz. The \textit{Leslie speaker} implementation\footnote{https://ccrma.stanford.edu/software/snd/snd/leslie.cms} is based on \cite{smith2002doppler} and we model each of the stereo channels. 

Finally, we also explore the capabilities of the model with nonlinear time-invariant audio effects with long temporal dependencies, such as \textit{compressors} and \textit{auto-wah}. We use the \textit{compressor} and \textit{multiband compressor} from \textit{SoX}\footnote{http://sox.sourceforge.net/} to process the electric guitar raw audio. The settings of the \textit{compressor} are as follows: attack time 10 ms, release time 100 ms, knee 1 dB, ratio 4:1 and threshold -40 dB. The \textit{multiband compressor} has 2 bands with a crossover frequency of $500$ Hz, attack time: $5$ ms and $625$ $\mu$s, decay time: $100$ ms and $12.5$ ms, knee: $0$ dB and $6$ dB, ratio: $3$:$1$ and $6$:$1$ and threshold: $-30$ dB and $-60$ dB.

Similarly, we use an \textit{auto-wah} implementation\footnote{https://github.com/lucieperrotta/ASP} with an envelope follower and a peak filter which center frequency modulates between $500$ Hz to $3$ kHz. 

For each time-varying task we use $624$ raw and effected notes and both the test and validation samples correspond to $5\%$ of this subset each. The recordings were downsampled to $16$ kHz and amplitude normalization was applied with exception to the time-invariant audio effects.

\subsection{Evaluation}

Two metrics were used when testing the models with the various test subsets. Since the mean absolute error depends on the amplitude of the output and target waveforms, before calculating this error, we normalize the energy of the target and the output and define it as the energy-normalized mean absolute error (\textit{mae}).

We also propose an objective metric which mimics human perception of amplitude and frequency modulation. The modulation spectrum uses time-frequency theory integrated with the psychoacoustics of modulation frequency perception, thus, providing long-term knowledge of temporal fluctuation patterns \cite{sukittanon2004modulation}. We propose the modulation spectrum euclidean distance (\textit{msed}), which is based on the audio features from \cite{mcdermott2011sound} and \cite{mckinney2003features} and is defined as follows: 

\begin{itemize}
    \item A Gammatone filter bank is applied to the target and output entire waveforms. In total we use $12$ filters, with center frequencies spaced logarithmically from $26$ Hz to $6950$ Hz.
    \item The envelope of each filter output is calculated via the magnitude of the Hilbert transform and downsampled to $400$ Hz.
    \item A Modulation filter bank is applied to each envelope. In total we use $12$ filters, with center frequencies spaced logarithmically from $0.5$ Hz to $100$ Hz.
    \item The Fast Fourier Transform (FFT) is calculated for each modulation filter output of each Gammatone filter. The energy is normalized by the DC value and summarized in the following bands: $0.5$-$4$ Hz, $4.5$-$10$ Hz, $10.5$-$20$ Hz and $20.5$-$100$ Hz.
    \item The \textit{msed} metric is the mean euclidean distance between the energy values at these $4$ bands.
\end{itemize}

\begin{figure*}[ht!]
\subfloat[]{
\begin{minipage}{0.5\textwidth}
\includegraphics[width=1.\linewidth,height=0.4\linewidth]{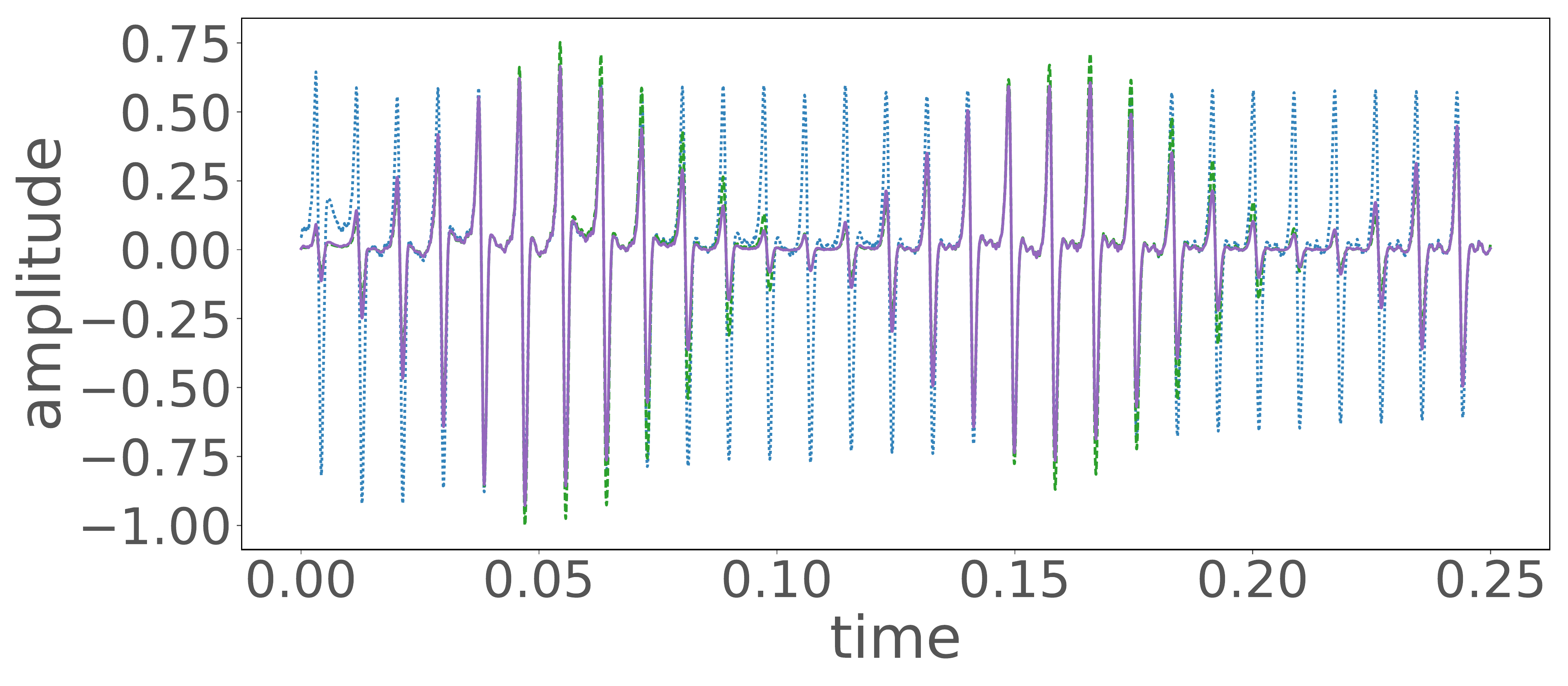}\label{fig:Tremolo-input_target_time}\\
\includegraphics[width=1.\linewidth, height=0.4\linewidth]{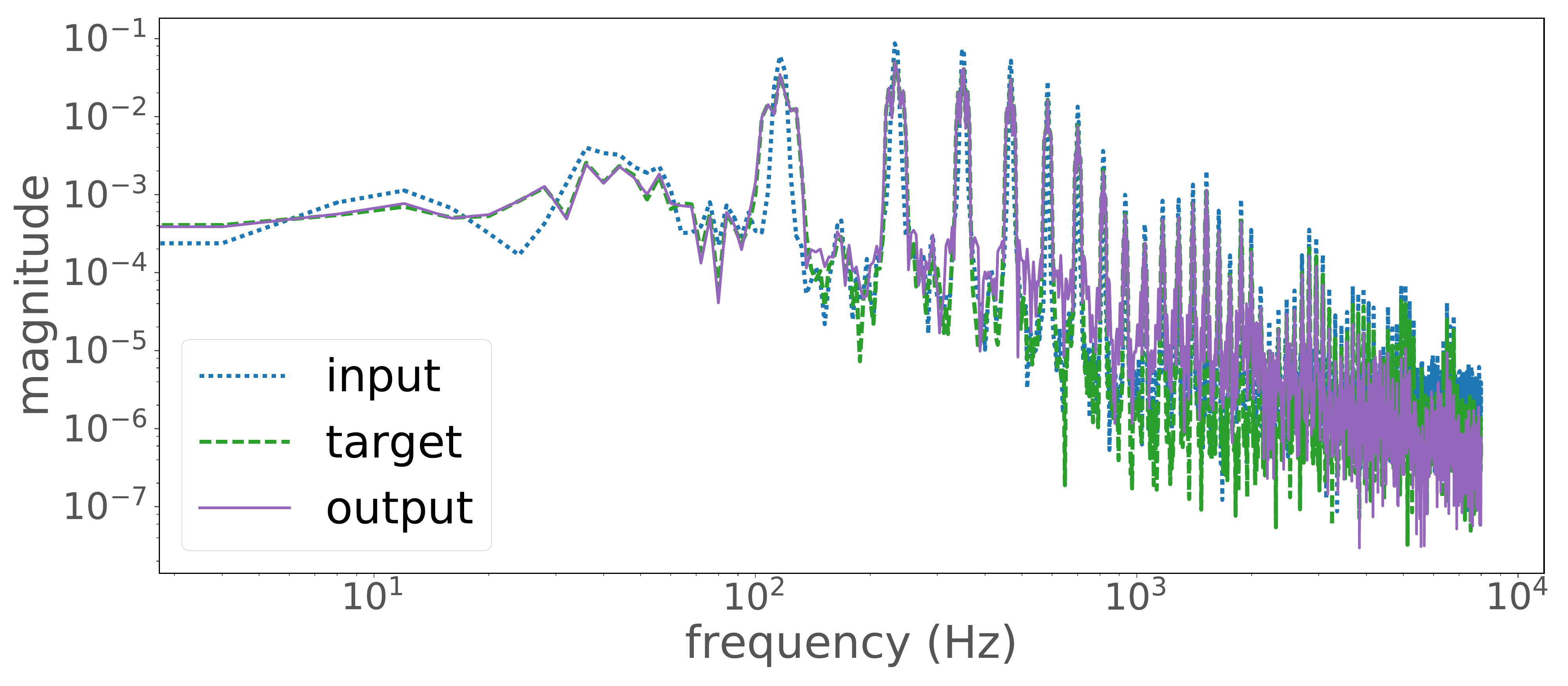}\label{fig:Tremolo-input_target_freq}
\end{minipage}}
\begin{minipage}{0.5\textwidth}
\subfloat[]{\includegraphics[width=.5\linewidth, height=0.35\linewidth]{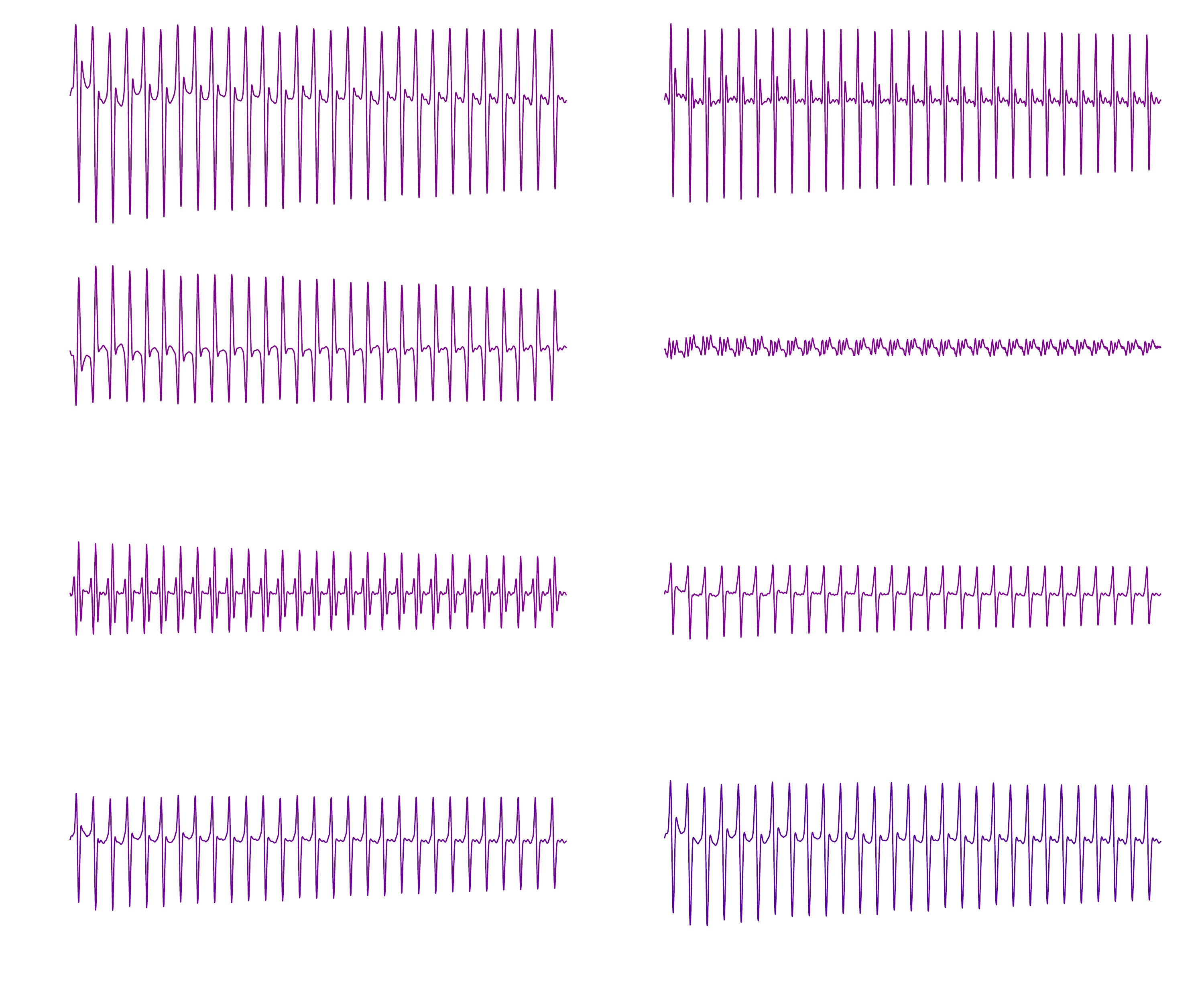}\label{fig:Tremolo-conv_output}}\hspace{0ex}
\subfloat[]{\includegraphics[width=.5\linewidth, height=0.35\linewidth]{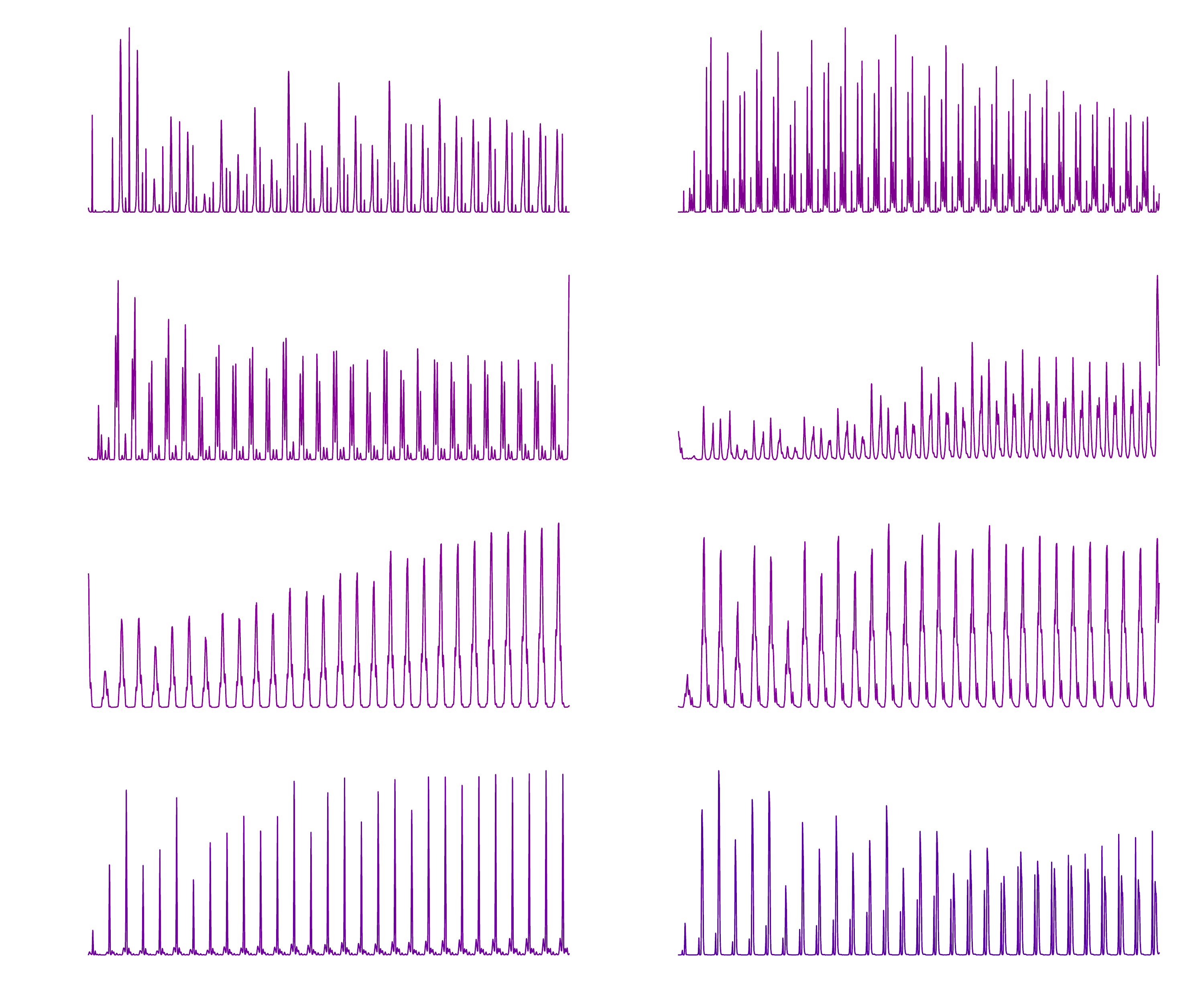}\label{fig:Tremolo-conv2}}\hspace{0ex}\\
\subfloat[]{\includegraphics[width=.5\linewidth, height=0.35\linewidth]{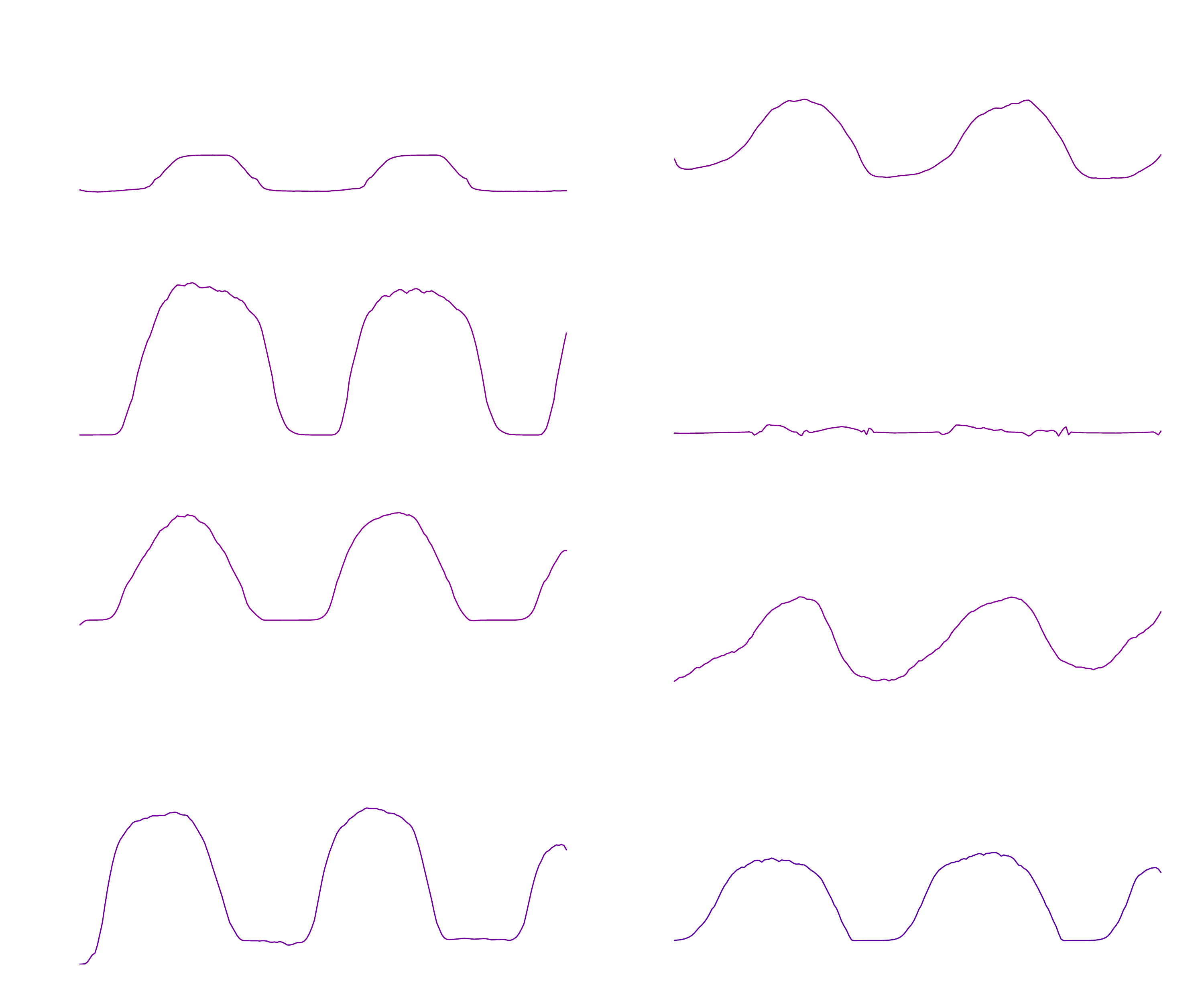}\label{fig:Tremolo-modulator}}\hspace{0ex}
\subfloat[]{\includegraphics[width=.5\linewidth, height=0.35\linewidth]{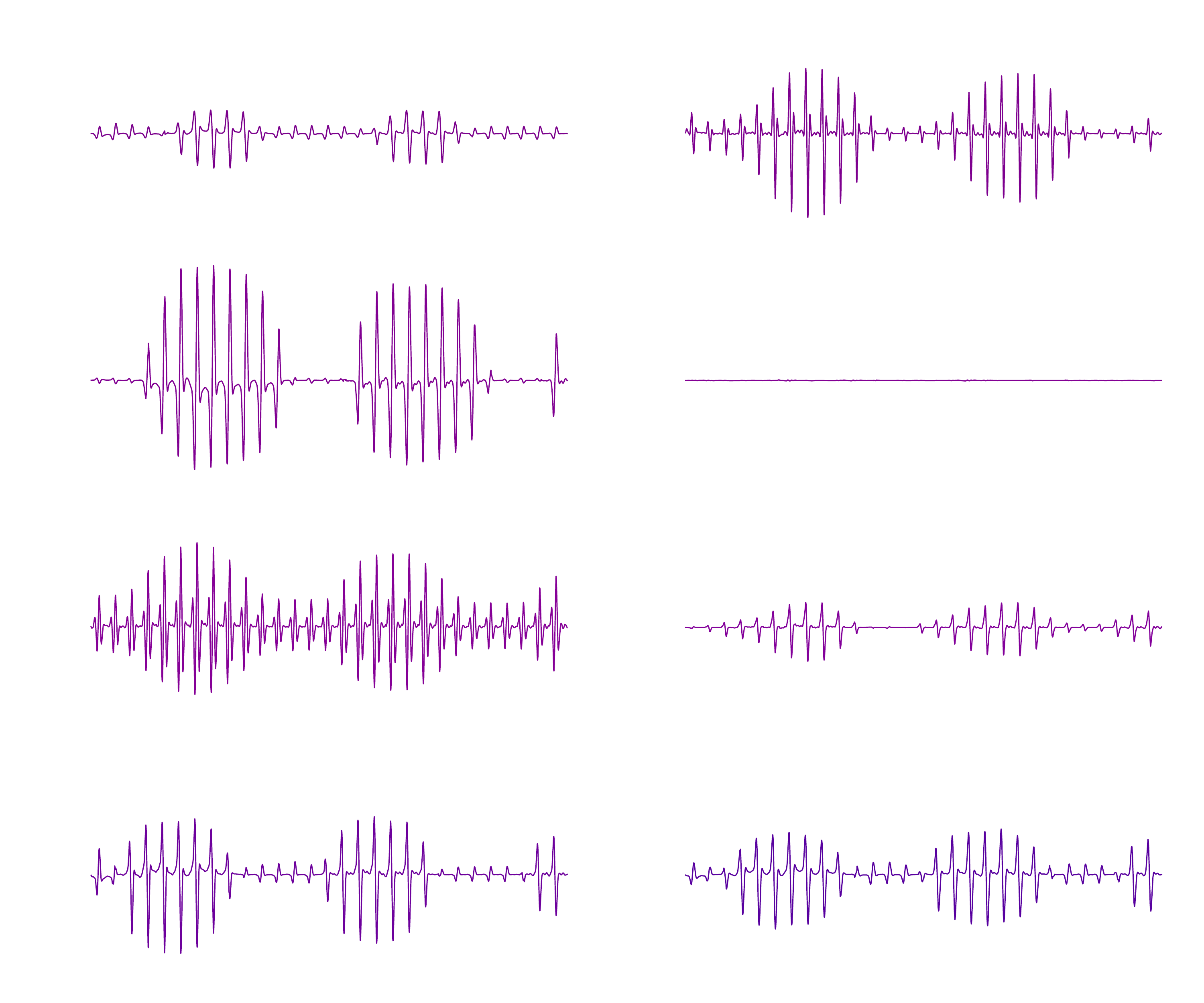}\label{fig:Tremolo-modulation}}\hspace{0ex}
\end{minipage}

\caption{\textit{Various internal plots for the test dataset of the \textbf{tremolo} modeling task. \ref{fig:Tremolo-input_target_time}) Input, target and output frames of $4096$ samples and their respective FFT magnitudes. \ref{fig:Tremolo-conv_output}) For the input frame $x(t)$, respective $8$ rows from $\boldsymbol{R}$. \ref{fig:Tremolo-conv2}) Following the filter bank architecture, respective $8$ rows from $\boldsymbol{X}_{2}$. \ref{fig:Tremolo-modulator}) From $\boldsymbol{\hat{Z}}$, corresponding $8$ modulator signals learned by the Bi-LSTM layer. \ref{fig:Tremolo-modulation}) In the same manner, $8$ rows from $\boldsymbol{\hat{X}}_{0}$, which is the input to the deconvolution layer prior to obtaining the output frame $\hat{y}(t)$. Vertical axes in \ref{fig:Tremolo-conv_output})-\ref{fig:Tremolo-modulation}) are unitless and horizontal axes correspond to time.}}
\label{fig:Tremolo-plots}
\end{figure*}

\section{Results \& Analysis}
\label{sec:analysis}

First, we explore the capabilities of Bi-LSTMs to learn long-term temporal dependencies. Fig. \ref{fig:mae-winlength} shows the $mae$ results of the test dataset for different input frame sizes and various linear time-varying tasks. The most optimal results are with an input size of $4096$ samples, since shorter frame sizes represent a higher error and $8192$ samples do not represent a significant improvement. Since the average modulation frequency in our tasks is $2$ Hz, for each input size we select a $k$ that covers one period of this modulator signal. Thus, for the rest of our experiments, we use an input size of $4096$ samples and $k=4$ for the number of past and subsequent frames.

The training procedures were performed for each type of time-varying and time-invariant audio effect. Then, the models were tested with samples from the test dataset and the audio results are available online\footnote{https://mchijmma.github.io/modeling-time-varying/}. Fig. \ref{fig:errors} shows the \textit{mae} and \textit{msed} for all the test subsets. To provide a reference, the mean \textit{mae} and \textit{msed} values between input and target waveforms are $0.15$ and $0.11$ respectively. It can be seen that the model performed well on each audio effect modeling task. Overall, the model achieved better results with amplitude modulation and time-varying filter audio effects, although delay-line based effects were also successfully modeled.

Fig. \ref{fig:Tremolo-plots}  visualizes the functioning of the model for the \textit{tremolo} task. It shows how the model processes the input frame $x(t)$ into the different frequency maps $\boldsymbol{X}_{1}$ and $\boldsymbol{X}_{2}$, learns a modulator signal $\boldsymbol{\hat{Z}}$, and applies the respective amplitude modulation. This linear time-varying audio effect is easy to interpret. For more complex nonlinear time-varying effects, a more in-depth analysis of the model is required.

\begin{figure*}[ht!]
\subfloat[]{
\begin{minipage}{.25\textwidth}
\includegraphics[width=1.\linewidth,height=1.2\linewidth]{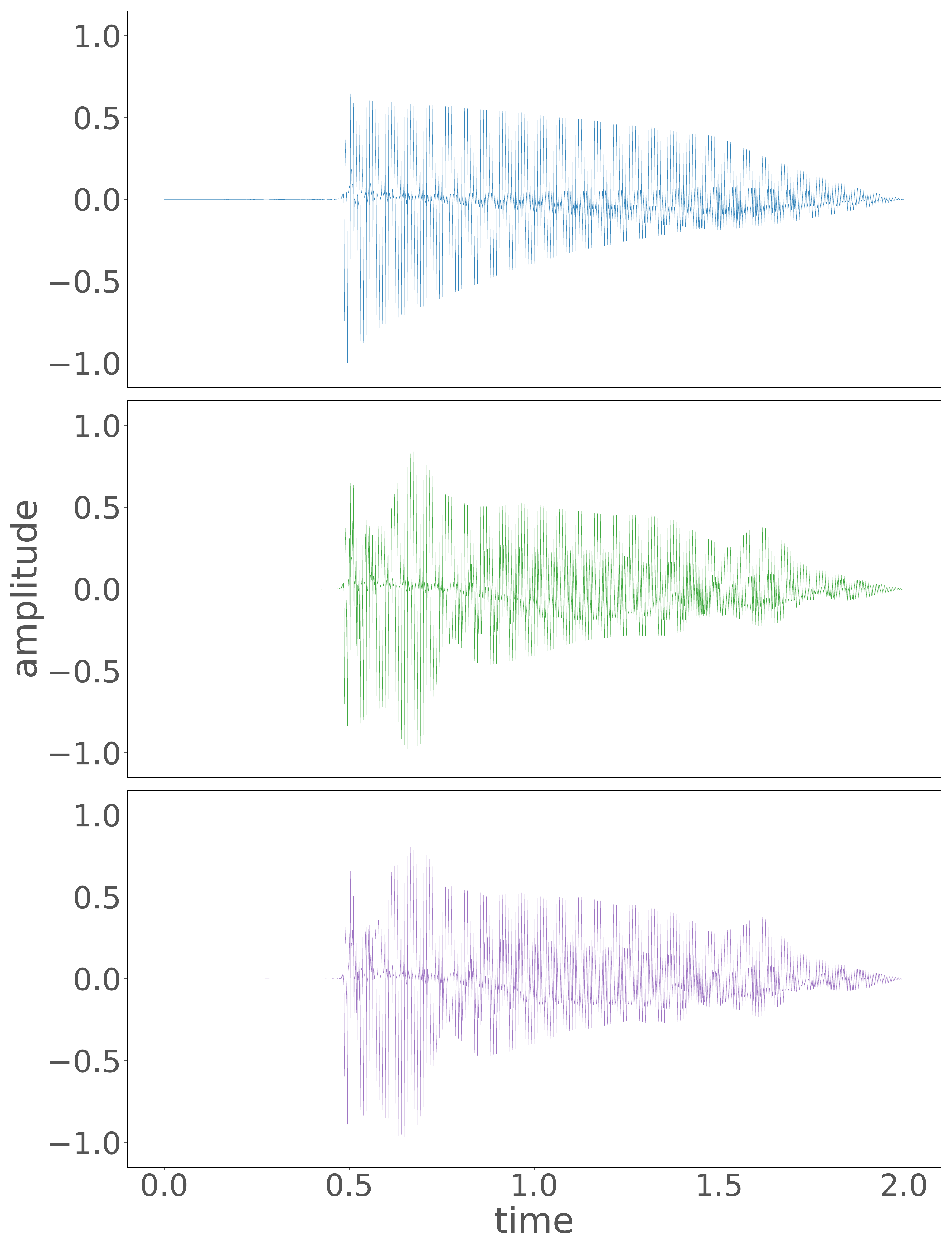}\label{fig:1-input}
\end{minipage}
\begin{minipage}{.25\textwidth}
\includegraphics[width=1.\linewidth,height=1.2\linewidth]{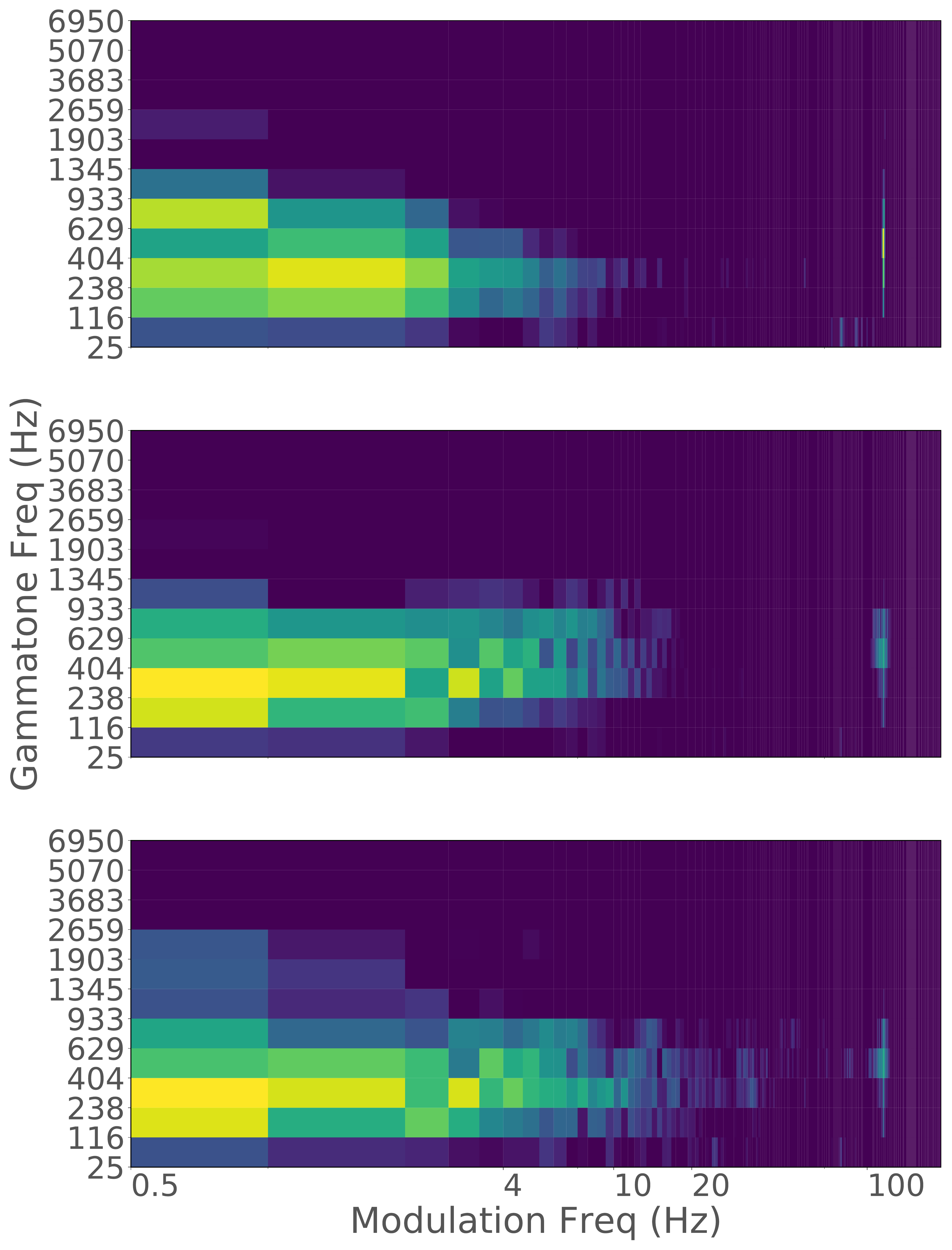}\label{fig:1-modulationSpectrum}
\end{minipage}}
\subfloat[]{
\begin{minipage}{.25\textwidth}
\includegraphics[width=1.\linewidth,height=1.2\linewidth]{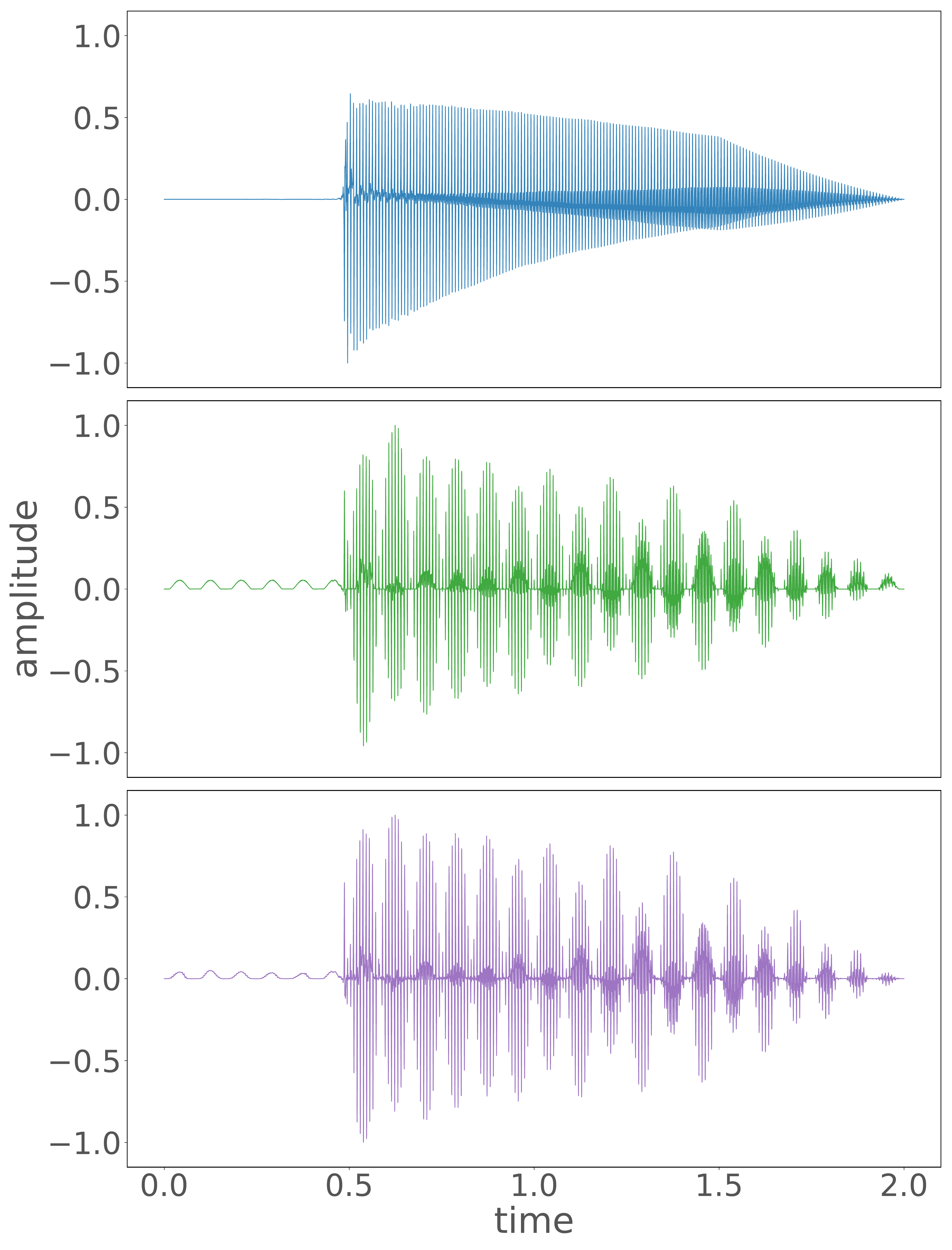}\label{fig:2-input}
\end{minipage}
\begin{minipage}{.25\textwidth}
\includegraphics[width=1.\linewidth,height=1.2\linewidth]{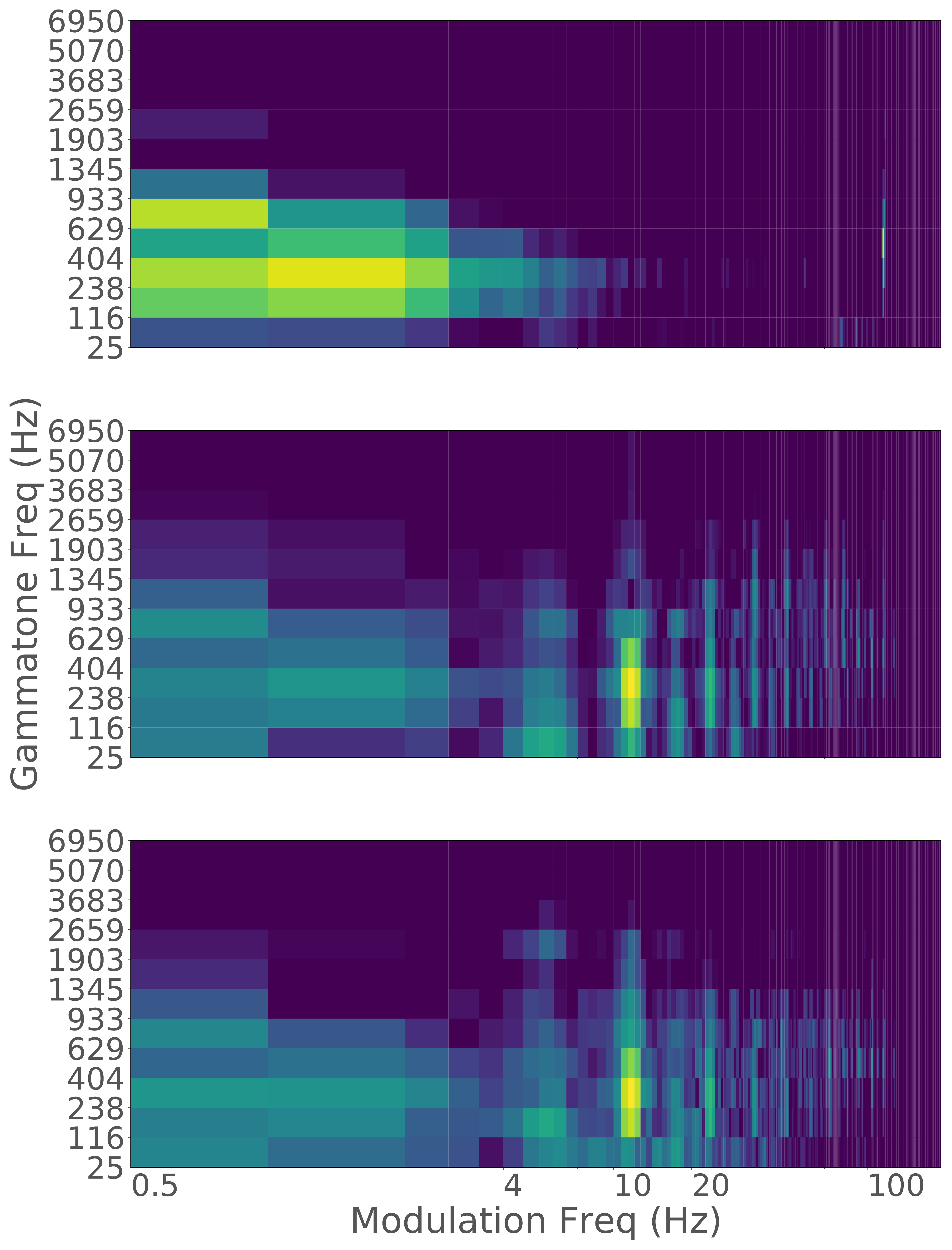}\label{fig:2-modulationSpectrum}
\end{minipage}}\\
\hspace{-4ex}
\subfloat[]{
\begin{minipage}{.25\textwidth}
\includegraphics[width=1.\linewidth,height=1.2\linewidth]{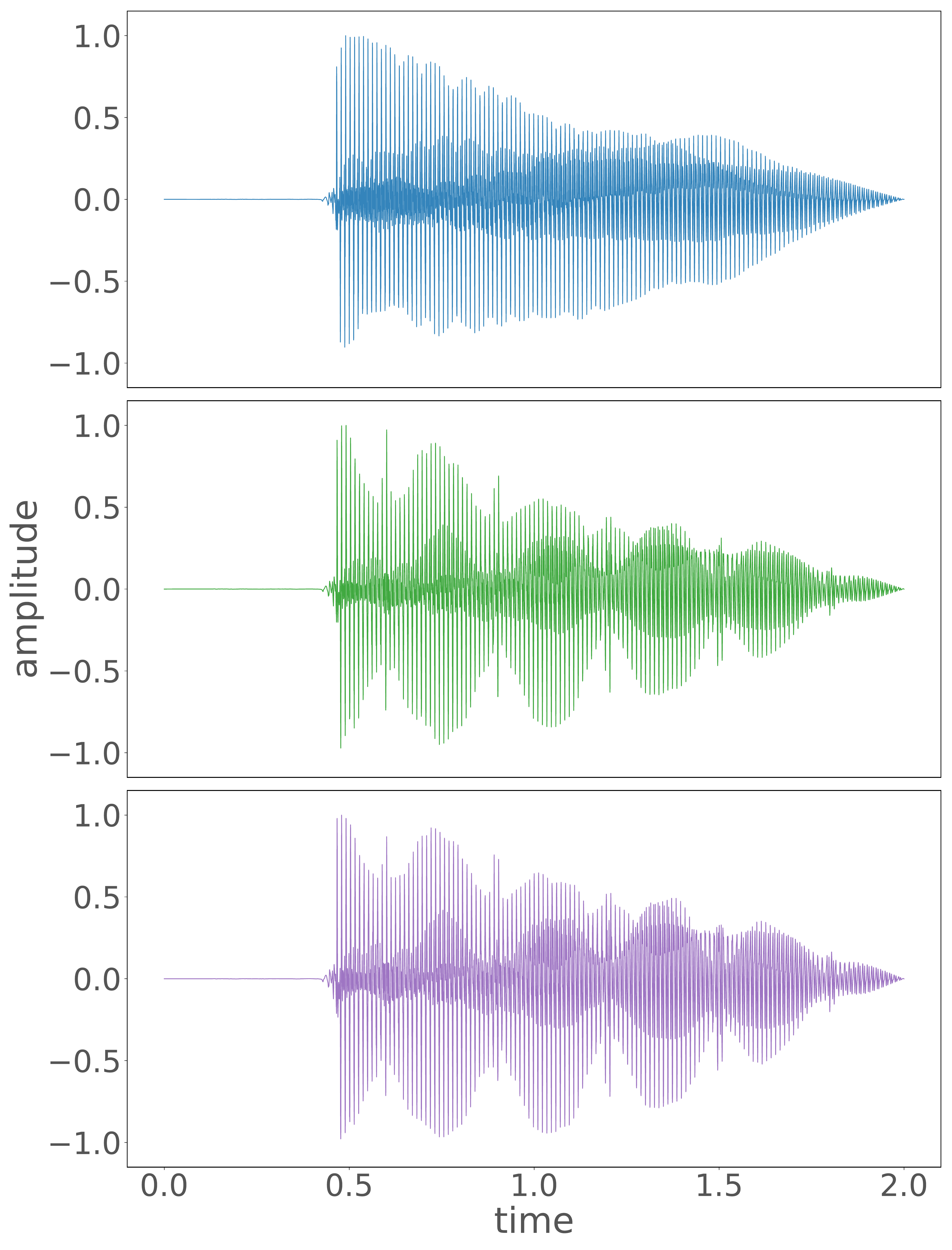}\label{fig:3-input}
\end{minipage}
\begin{minipage}{.25\textwidth}
\includegraphics[width=1.\linewidth,height=1.2\linewidth]{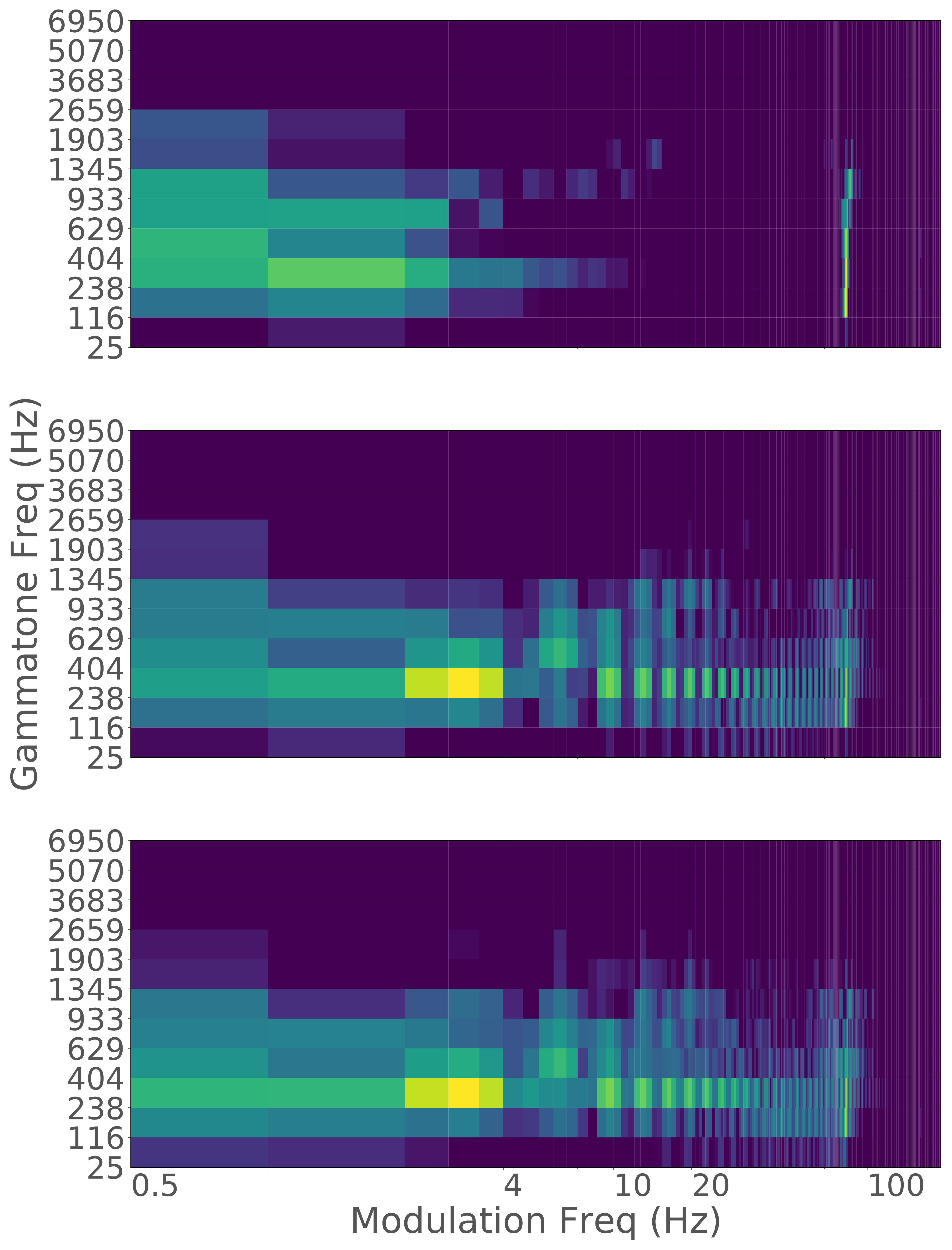}\label{fig:3-modulationSpectrum}
\end{minipage}}
\subfloat[]{
\begin{minipage}{.25\textwidth}
\includegraphics[width=1.\linewidth,height=1.2\linewidth]{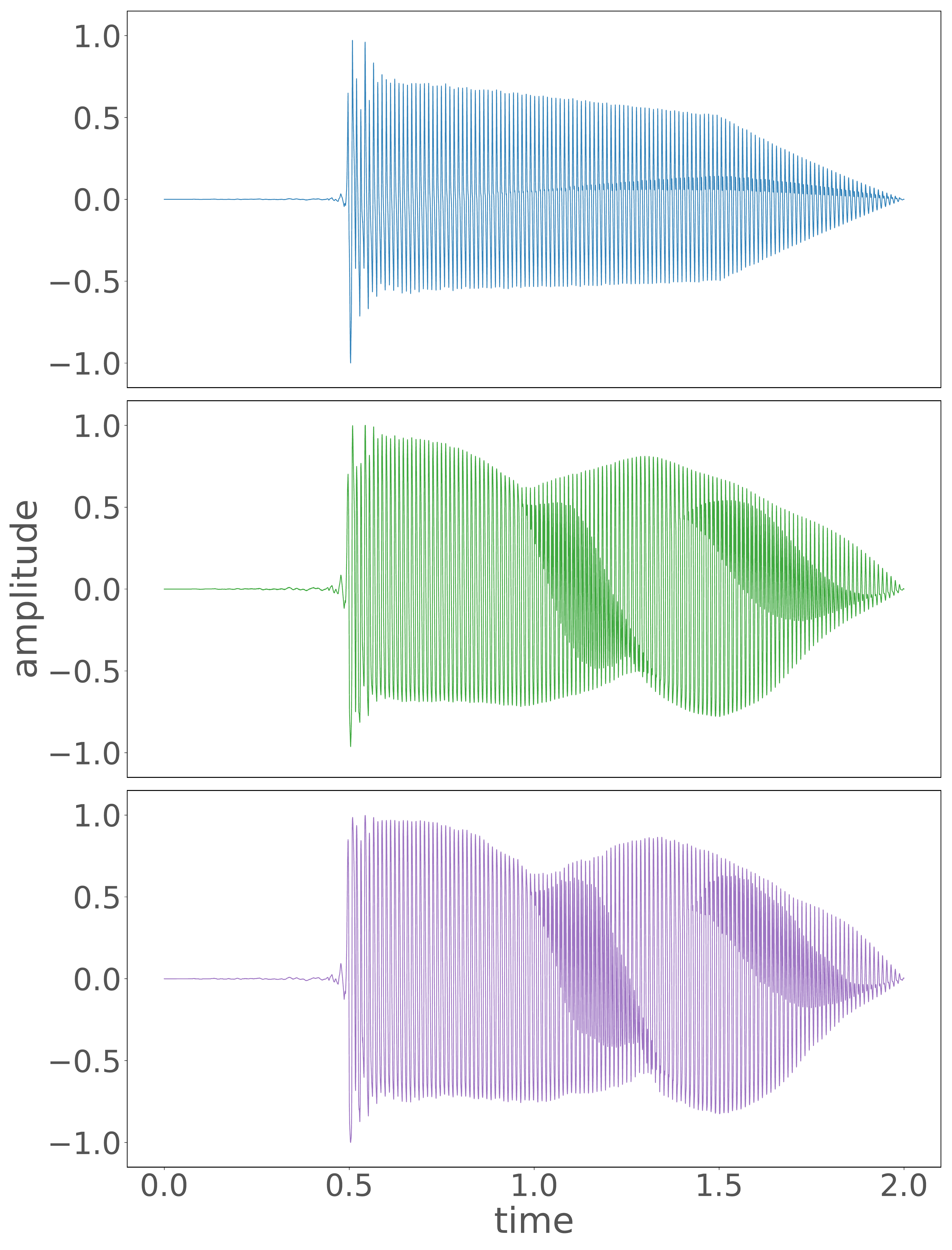}\label{fig:4-input}
\end{minipage}
\begin{minipage}{.25\textwidth}
\includegraphics[width=1.\linewidth,height=1.2\linewidth]{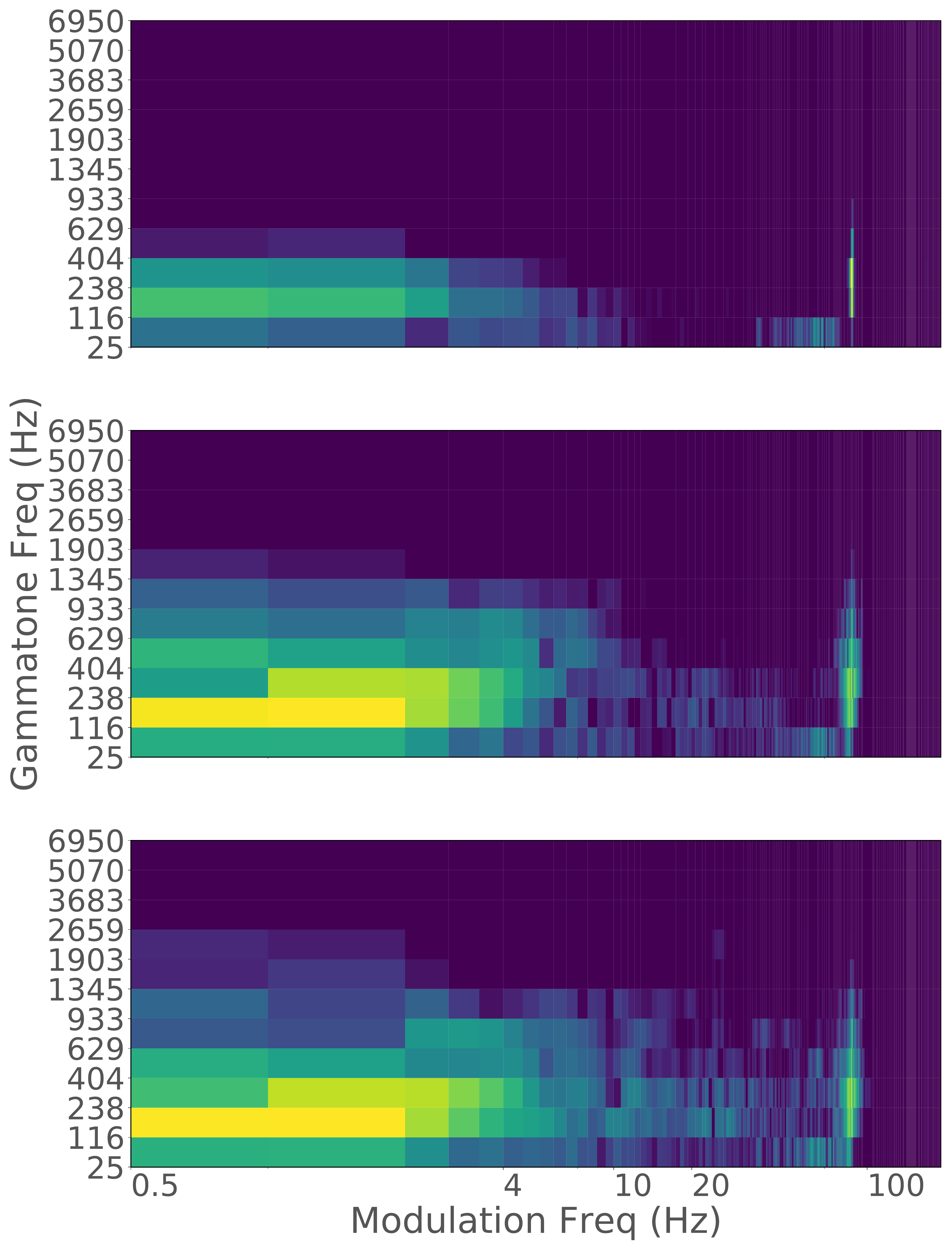}\label{fig:4-modulationSpectrum}
\end{minipage}}
\caption {\textit{Results with the test dataset for the following time-varying tasks: \ref{fig:1-modulationSpectrum}) \textbf{chorus}, \ref{fig:2-modulationSpectrum}) \textbf{ring modulator}, \ref{fig:3-modulationSpectrum}) \textbf{Leslie speaker} (right channel) and \ref{fig:4-modulationSpectrum}) \textbf{phaser-overdrive}. For each subfigure and from top to bottom: input, target and output waveforms and respective modulation spectrum plots.}}
\label{fig:results}
\end{figure*}

For selected linear and nonlinear time-varying tasks, Fig. \ref{fig:results} shows the input, target, and output waveforms together with their respective modulation spectrum. In the time-domain, it is evident that the model is matching the target waveform. From the modulation spectrum it is noticeable that the model introduces different modulation energies into the output which were not present in the input and which closely match those of the respective targets. 

The task becomes more challenging when a nonlinearity is added to a linear time-varying transformation. Fig. \ref{fig:4-input} depicts results for the \textit{phaser-overdrive} task. Given the large overdrive gain the resulting audio has a lower-frequency modulation. It can be seen that the model introduces modulations as low as $0.5$ Hz. But the waveform is not as smooth as the target, hence the larger $mae$ values. Although the $mae$ increased, the model does not significantly reduce performance and is able to match the combination of nonlinear and modulation based audio effects. 

Much more complicated time-varying tasks, such as the \textit{ring modulator} and \textit{Leslie speaker} virtual analog implementations were also successfully modeled. This represents a significant result, since these implementations include nonlinear modulation; \textit{ring modulator}, or varying delay lines together with artificial reverberation and Doppler effect simulation; and the \textit{Leslie speaker}.

Lastly, the model is also able to perform linear and nonlinear time-invariant modeling. The long temporal dependencies of an envelope driven \textit{auto-wah}, \textit{compressor} and \textit{multiband compressor} are succesfully modeled. Furthermore, in the latter case, the crossover filters are also matched. The \textit{msed} may not be relevant for these effects, but the low \textit{mae} values represent that the model also performs well here.

\section{Conclusion}
\label{sec:conclusion}

In this work, we introduced a general-purpose deep learning architecture for modeling audio effects with long temporal dependencies. Using raw audio and a given time-varying task, we explored the capabilities of end-to-end deep neural networks to learn low-frequency modulations and to process the audio accordingly.
The model was able to match linear and nonlinear time-varying audio effects, time-varying virtual analog implementations and time-invariant audio effects with long-term memory. 

Other white-box or gray-box modeling methods suitable for these time-varying tasks would require expert knowledge such as specific circuit analysis and discretization techniques. Moreover, these methods can not easily be extended to other time-varying tasks, and assumptions are often made regarding the nonlinear behavior of certain components. To the best of our knowledge, this work represents the first black-box modeling method for linear and nonlinear, time-varying and time-invariant audio effects. It makes less assumptions about the audio processor target and represents an improvement of the state-of-the-art in audio effects modeling. 

Using a small amount of training examples we showed the model matching \textit{chorus}, \textit{flanger}, \textit{phaser}, \textit{tremo\-lo}, \textit{vibrato}, \textit{auto-wah}, \textit{ring modulator}, \textit{Leslie speaker} and \textit{compressors}. We proposed an objective perceptual metric to measure the performance of the model. The metric is based on the euclidean distance between the frequency bands of interest within the modulation spectrum. We demonstrated that the model processes the input audio by applying different modulations which closely match with those of the time-varying target. 

Perceptually, most output waveforms are indistinguishable from their target counterparts, although there are minor discrepancies at the highest frequencies and noise level. This could be improved by using more convolution filters, which means a higher resolution in the filter bank structures \cite{martinez2019modeling}. Moreover, as shown in \cite{martinez2018end}, a cost function based on time and frequency can be used to improve this frequency related issue, though listening tests may be required.

The generalization can also be studied more thoroughly, since the model learns to apply the specific transformation to the audio of a specific musical instrument, such as the electric guitar or the bass guitar. In addition, since the model strives to learn long temporal dependencies with shorter input size frames, and also needs past and subsequent frames, more research is needed on how to adapt this architecture to real-time implementations.

Real-time applications would benefit significantly from the exploration of recurrent neural networks to model transformations that involve long-term memory without resorting to large input frame sizes and the need for past and future context frames. Although the model was able to match the artificial reverberation of the \textit{Leslie speaker} implementation, a thorough exploration of reverberation modeling is needed, such as plate, spring or convolution reverberation. In addition, since the model is learning a static representation of the audio effect, ways of devising a parametric model could also be explored. Finally, applications beyond virtual analog can be investigated, for example, in the field of automatic mixing the model could be trained to learn a generalization from mixing practices.

\newpage

\section{Acknowledgments}
The Titan Xp GPU used for this research was donated by the NVIDIA Corporation. EB is supported by a RAEng Research Fellowship (RF/128).

\footnotesize

\bibliographystyle{IEEEbib}
\bibliography{bib_refs} 
\end{document}